\documentclass[sigconf]{acmart} 
\AtBeginDocument{%
  }

\copyrightyear{2026}
\acmYear{2026}
\setcopyright{cc}
\setcctype{by}
\acmConference[AISec '26]{19th Workshop on Artificial Intelligence and Security}{November 15--19, 2026}{The Hague, Netherlands}
\acmBooktitle{19th Workshop on Artificial Intelligence and Security (AISec '26), November 15--19, 2026, The Hague, Netherlands}
\acmDOI{10.1145/3847352.3848112}
\acmISBN{979-8-4007-3031-3/2026/11}

\usepackage{booktabs}
\usepackage{tikz}
\usetikzlibrary{positioning,fit,calc}
\usepackage{multirow}
\usepackage{graphicx}
\usepackage{cleveref}
\usepackage{enumitem}
\usepackage{siunitx}
\usepackage{subcaption}
\usepackage{framed}

\begin{document}

\title{TRIAGE-CPG: A Cross-Model Study of Budgeted Static Malware Triage with Large Language Models}



\author{Paul-Andrei Sava}
\orcid{0000-0003-3027-9500}
\email{paul-andrei.sava@aisec.fraunhofer.de}

\author{Alexander Küchler}
\orcid{0000-0001-6633-460X}
\email{alexander.kuechler@aisec.fraunhofer.de}

\author{Daniel Kowatsch}
\orcid{0009-0000-9755-1594}
\email{daniel.kowatsch@aisec.fraunhofer.de}
\affiliation{%
  \institution{Fraunhofer AISEC}
  \city{Garching}
  \country{Germany}
}

\author{Bernhard Grill}
\orcid{0009-0006-8892-2465}
\email{bgrill@google.com}

\author{Jennifer Pullman}
\orcid{0009-0005-6861-6885}
\email{jpullman@google.com}

\author{Yanick Fratantonio}
\orcid{0009-0008-3676-9117}
\email{yanickf@google.com}

\author{Luca Invernizzi}
\orcid{0000-0001-8420-0760}
\email{invernizzi@google.com}

\affiliation{%
 \institution{Google}
 \city{Zurich}
  \country{Switzerland}
}

%
%
%
%
%
%

\renewcommand{\shortauthors}{Sava et al.}

\begin{abstract} 
When triaging unknown binaries, security analysts must quickly decide which samples to clear as benign, flag as malicious, or escalate for deeper analysis.
Large Language Models (LLMs) could support these decisions by interpreting decompiled code, but a binary can contain hundreds or thousands of functions, making exhaustive LLM analysis impractical.
We study how current LLMs behave when static triage is formulated as budgeted evidence acquisition: which functions they inspect and how they trade off benign clearance, malware detection, and escalation.

To enable a controlled comparison, we implement TRIAGE-CPG, a fixed pipeline over a code property graph (CPG) of a decompiled binary.
An LLM-free heuristic selects seed functions, an agent expands the resulting evidence dossier under a fixed budget, and a reporting stage produces a benign, suspicious, or malicious verdict together with a behavioral report.
We evaluate eight closed-source and open-weight models on \num{1000} Windows PE binaries, using a 600-sample test split and the same zero-shot protocol for every model.
We assess the ternary verdicts under both a strict binary reading and triage balanced accuracy (tBA), which credits malware that is flagged or escalated and benign samples only when cleared.

After analyzing about 40 of a mean \num{1000} functions per binary, the highest-scoring model reaches a strict balanced accuracy of 0.805 and a tBA of 0.855, flagging 65\% of malware, escalating another 12\%, and clearing 94\% of benign samples.
The models implement different triage policies: high-specificity models produce small, malware-rich escalation queues, whereas recall-oriented models escalate or flag much more broadly.
Both closed-source and open-weight models appear across this spectrum.
Our results identify no universal deployment configuration: model choice itself determines the clear/flag/escalate trade-off under limited static evidence.
\end{abstract}




\maketitle

\section{Introduction}\label{sec:intro}

Threat hunters face large volumes of unknown binaries and must decide quickly and safely which ones deserve scarce human attention \cite{baileyAutomatedClassificationAnalysis2007, bayerScalableBehaviorBasedMalware}.
Dynamic analysis is the de facto standard \cite{egeleSurveyAutomatedDynamic2008}, but it is costly \cite{botacinWhatMalwareAnalysts2024}, requires a secure environment in which it can execute the sample \cite{grazianoNeedlesHaystackMining2015}, and can still be evaded by environment-aware \cite{kiratBareCloudBaremetalAnalysisbased, miramirkhaniSpotlessSandboxesEvading2017} or stalling malware \cite{kolbitschPowerProcrastinationDetection2011, DoesEverySecond}. 
On the other hand, static analysis avoids execution and can in principle inspect the entire program, but may be impeded by packing and obfuscation \cite{moserLimitsStaticAnalysis2007}. 
It also carries a scaling cost: a single binary can decompile into hundreds or thousands of functions \cite{marcelliHowMachineLearning2022}, far more than an analyst or a fixed LLM context window can read.
These constraints motivate a bounded first pass that clears likely benign samples, flags samples with malicious evidence, and escalates cases that cannot be resolved from available static evidence.

The consequences of these decisions are asymmetric.
Within an intake workflow, clearing malware as benign is difficult to recover from because the sample leaves the analysis queue.
Escalating a benign sample does consume downstream analysis resources, but the error can still be corrected.
A triage procedure should therefore expose uncertainty rather than forcing every sample into a committed benign or malicious verdict.
We represent this uncertainty through a third verdict, \emph{suspicious}, which routes the sample to deeper analysis.

Large Language Models (LLMs) are a plausible tool for this setting because they can interpret decompiled code and use external tools to acquire additional context \cite{yangSWEagentAgentcomputerInterfaces2024, yao2023react}. 
Code property graphs (CPGs) provide a corresponding navigation substrate by combining call relationships with control- and data-flow information in a queryable program representation \cite{yamaguchiModelingDiscoveringVulnerabilities2014, weissLanguageIndependentAnalysisPlatform2022}.
However, these capabilities do not remove the underlying scale constraint.
Whole binaries remain too large or costly to inspect exhaustively \cite{wangLargeLanguageModels2026}, many recovered functions contain boilerplate or library code, and unbounded agent exploration is expensive and difficult to compare reproducibly~\cite{mohammadiEvaluationBenchmarkingLLM2025}.
This raises the empirical question we study: how do current LLMs behave when they must triage a binary from a small, graph-selected subset of its functions, and how does model choice affect the balance between clearance, malware detection, and escalation?

To investigate this question, we introduce TRIAGE-CPG, a fixed, graph-guided pipeline for static triage of Windows PE binaries.
The system receives a CPG obtained from a decompiled binary together with its PE metadata and operates in three stages.
First, an LLM-free heuristic ranks the recovered functions and analyzes the highest-ranked ones to initialize an evidence dossier.
Second, an agent expands this dossier by following call-graph leads under a fixed exploration budget. 
Finally, a reporting agent reviews the frozen dossier and returns a benign, suspicious, or malicious verdict together with a behavioral report. 
TRIAGE-CPG therefore provides a common evidence interface through which different models can be compared under the same constraints.

We evaluate the ternary verdicts under two complementary readings: a \emph{strict} reading, in which only a committed malicious verdict counts as positive, and a \emph{triage} reading, in which malware counts as handled when flagged or escalated while a benign sample counts only when cleared.
We call balanced accuracy under the latter \emph{triage balanced accuracy} (tBA) and report it alongside the strict reading and explicit escalation-queue measurements.

We compare eight closed-source and open-weight models on a locked test set of 600 Windows PE binaries under one uniform, zero-shot protocol.
The fixed pipeline is intended to isolate differences in model behavior rather than to optimize a separate configuration for each model.
Under this protocol, the highest-scoring model reaches a tBA of 0.855 while inspecting approximately 40 recovered functions per binary.
It flags or escalates 77\% of malware and clears 94\% of benign samples.

The models also do not converge on a common triage policy.
High-specificity models clear most benign samples and produce small, malware-rich escalation queues, but issue fewer committed malware verdicts.
Recall-oriented models flag or escalate more broadly, catching more malware at the cost of routing substantially more benign samples to further analysis.
Both closed-source and open-weight models appear across this spectrum, so model choice is not just an implementation detail: under otherwise identical conditions, it determines the operating point of the triage workflow.

\paragraph{Our Contributions}
We make the following contributions:
\begin{itemize}
    \item We formulate static LLM malware triage as budgeted, escalation-oriented evidence acquisition. We define and evaluate \textit{triage balanced accuracy} (tBA), the balanced accuracy of the clear-versus-not-clear decision, and report it together with a strict verdict reading and explicit escalation-queue metrics.
    \item We introduce TRIAGE-CPG as a controlled evidence interface for studying this setting. The pipeline combines LLM-free function prioritization, bounded CPG exploration, and ternary adjudication while holding the available evidence and action budgets fixed across models.
    \item We conduct a fixed-protocol comparison of eight closed-source and open-weight models on a locked 600-binary test split. We find that models occupy substantially different recall--specificity and escalation operating points, showing that model choice itself determines the resulting triage policy.
\end{itemize}

\section{Approach}
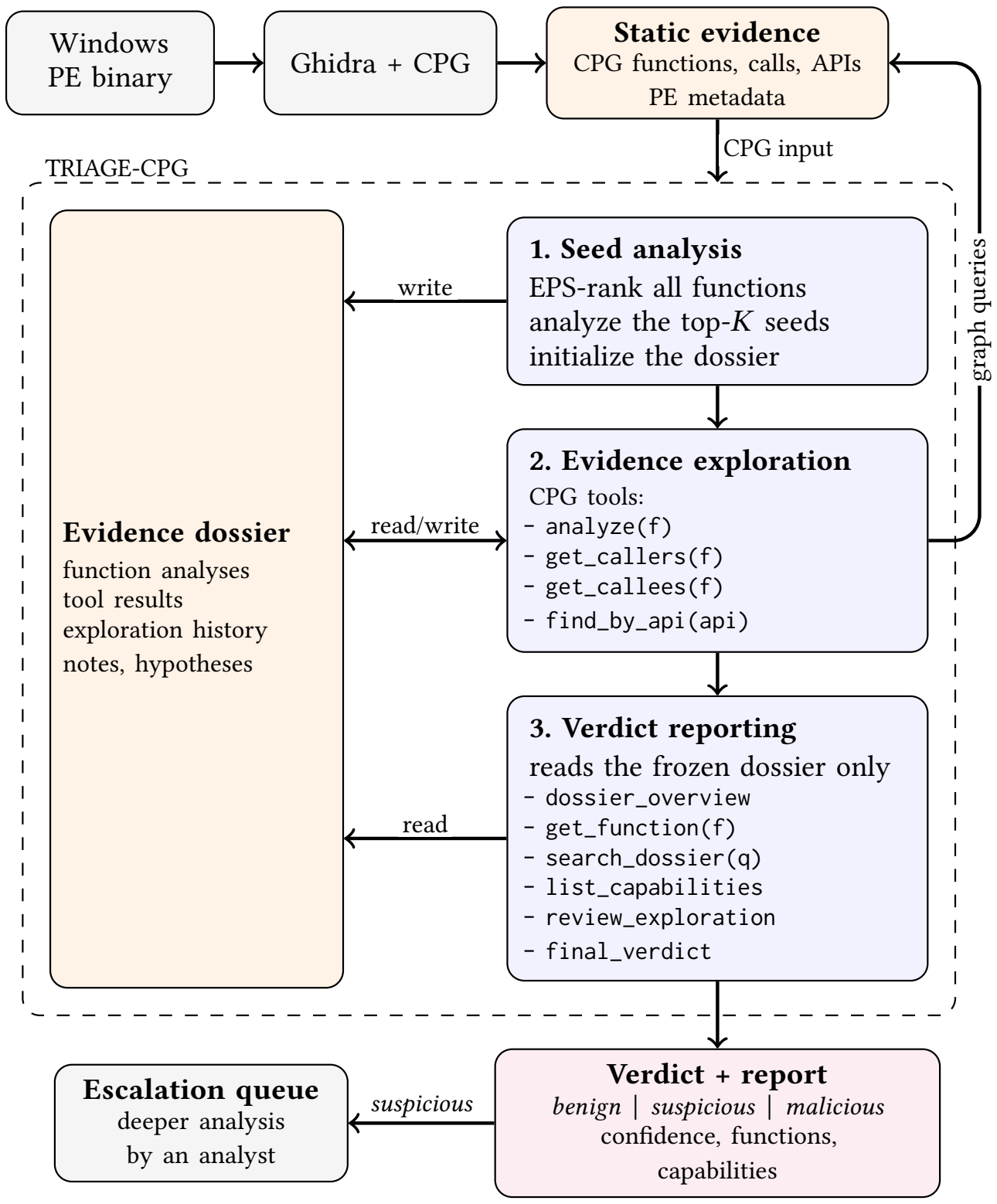
\begin{figure}[t]
\centering
\resizebox{\columnwidth}{!}{%
\begin{tikzpicture}[
    font=\footnotesize,
    box/.style={draw, rounded corners, align=center, minimum height=0.85cm, fill=gray!8},
    store/.style={draw, rounded corners, align=center, fill=orange!10},
    stage/.style={draw, rounded corners, align=left, text width=3.1cm,
                  inner xsep=0.18cm, inner ysep=0.16cm, fill=blue!5},
    outbox/.style={draw, rounded corners, align=center, fill=purple!7},
    arrow/.style={->, thick},
    rw/.style={<->, thick},
    lab/.style={fill=white, inner sep=1pt, font=\scriptsize}
]

\node[box, text width=1.5cm] (binary) {Windows PE binary};
\node[box, right=0.4cm of binary, text width=1.7cm] (frontend) {Ghidra + CPG};
\node[store, right=0.4cm of frontend, text width=2.6cm] (cpg)
      {\textbf{Static evidence}\\{\scriptsize CPG functions, calls, APIs\\PE metadata}};

\node[stage, below=0.8cm of cpg] (seed) {
\textbf{1.\ Seed analysis}\\[0.15em]
EPS-rank all functions\\
analyze the top-$K$ seeds\\
initialize the dossier
};
\node[stage, below=0.35cm of seed] (explore) {
\textbf{2.\ Evidence exploration}\\[0.15em]
{\scriptsize CPG tools:\\
\texttt{- analyze(f)}\\
\texttt{- get\_callers(f)}\\
\texttt{- get\_callees(f)}\\
\texttt{- find\_by\_api(api)}
}
};
\node[stage, below=0.35cm of explore] (report) {
\textbf{3.\ Verdict reporting}\\[0.15em]
reads the frozen dossier only\\
{\scriptsize\texttt{- dossier\_overview}\\
\texttt{- get\_function(f)}\\
\texttt{- search\_dossier(q)}\\
\texttt{- list\_capabilities}\\
\texttt{- review\_exploration}\\
\texttt{- final\_verdict}}
};

\node[store, align=left, text width=2.2cm, minimum height=6.4cm] (dossier)
    at ($(seed.north west)!0.5!(report.south west)+(-2.55,0)$)
    {\textbf{Evidence dossier}\\[0.2em]\scriptsize function analyses\\tool results\\exploration history\\notes, hypotheses};

\node[draw, dashed, rounded corners, inner sep=0.22cm,
      fit=(seed)(explore)(report)(dossier)] (TRIAGE-CPG) {};
\node[font=\bfseries, lab, anchor=south west]
    at ([xshift=0.15cm]TRIAGE-CPG.north west) {TRIAGE-CPG};

\draw[arrow] (binary) -- (frontend);
\draw[arrow] (frontend) -- (cpg);
\draw[arrow] (cpg.south) -- node[right, lab] {CPG input}
    (cpg.south |- TRIAGE-CPG.north);
\draw[arrow] (seed) -- (explore);
\draw[arrow] (explore) -- (report);

\draw[arrow, rounded corners=6pt]
    (explore.east) -- ++(0.42,0) |- (cpg.east);
\node[lab, rotate=90] at ($(explore.east)+(0.42,1.9)$) {graph queries};

\draw[arrow] (seed.west) -- node[above, lab] {write} (dossier.east |- seed.west);
\draw[rw] (explore.west) -- node[above, lab] {read/write} (dossier.east |- explore.west);
\draw[arrow] (report.west) -- node[above, lab] {read} (dossier.east |- report.west);

\node[outbox, below=0.55cm of report, text width=3.45cm] (output)
    {\textbf{Verdict + report}\\{\scriptsize\textit{benign} $\mid$ \textit{suspicious} $\mid$ \textit{malicious}\\
    confidence, functions, \\  capabilities}};
\node[box, left=1.2cm of output, text width=2.2cm] (queue)
    {\textbf{Escalation queue}\\{\scriptsize deeper analysis\\by an analyst}};
\draw[arrow] (report) -- (output);
\draw[arrow] (output) -- node[above, lab] {\textit{suspicious}} (queue);

\end{tikzpicture}%
}
\Description{Diagram of the TRIAGE-CPG workflow. A Windows PE binary is transformed by Ghidra and a code property graph into static evidence containing CPG functions, calls, APIs, and PE metadata. 
This evidence feeds a dashed TRIAGE-CPG box. Seed analysis initializes an evidence dossier; evidence exploration queries the CPG and reads from and writes to the dossier; verdict reporting reads the frozen dossier. The system produces a benign, suspicious, or malicious verdict and a report. 
Suspicious samples are sent to an escalation queue for deeper analyst analysis.}
\caption{TRIAGE-CPG pipeline. The CPG and PE metadata of a decompiled Windows PE binary provide the static evidence. EPS selects seed functions to initialize a dossier; an exploration agent may expand it under a fixed function-read budget. A reporting agent reads the frozen dossier and returns a ternary verdict and behavioral report; suspicious samples are escalated.
}

\label{fig:system-overview}
\end{figure}

TRIAGE-CPG processes each Windows PE binary through a common evidence interface in two phases.
A preprocessing step decompiles the binary with Ghidra\footnote{\url{https://github.com/nationalsecurityagency/ghidra}}, constructs a code property graph (CPG), and extracts PE metadata.
The resulting static evidence is then passed to a three-stage LLM workflow, shown in \Cref{fig:system-overview}: an LLM-free heuristic selects seed functions and initializes an evidence dossier, an exploration agent follows leads through the graph under a fixed budget, and a reporting agent reviews the collected evidence and produces a ternary verdict and behavioral report.

The reported evidence budget covers model actions used to analyze functions or inspect the dossier.
It does not include decompilation or CPG construction and therefore measures model-facing evidence use rather than end-to-end analysis time.
\Cref{sec:eps} describes the seeding heuristic, while Sections~\ref{sec:exploration} and~\ref{sec:verdict} present the exploration and reporting stages.

\subsection{CPG-based Evidence Representation}\label{sec:cpg}
The preprocessing stage constructs the CPG using an open-source, language-independent implementation \cite{weissLanguageIndependentAnalysisPlatform2022}.
The CPG represents syntax, control flow, data flow, call relationships, and program dependence in a single queryable graph.
This representation is useful for LLM-assisted analysis because it supports the extraction of bounded program context even when the available code is incomplete, as is common in decompiled binaries \cite{lekssaysLLMxCPGContextAwareVulnerability, weissLanguageIndependentAnalysisPlatform2022}.
Furthermore, it serves as abstract representation for different types of malware.

Although the constructed CPG retains control-flow, data-flow, call, and program-dependence edges, the workflow evaluated in this paper uses only its function-level call-graph projection.
Future extensions, however, can make use of the more powerful sub-graphs to enhance the analysis capabilities. 
E.g., program slicing can reduce the LLM input to the required context only, or the built-in data-flow analysis can be extended to taint-style analysis providing additional signals to identify relevant code regions.

Each node in the CPG represents a recovered function together with its decompiled body.
Internal edges represent calls between recovered functions, while calls without a matching in-binary function are represented as external API references.
Callers and callees provide the graph relations exposed to the exploration agent.
The function is also the unit charged to the evidence budget: analyzing a function reveals its complete decompiled body.
The remaining CPG edge types are retained in the serialized representation but are not queried by the evaluated workflow and therefore do not affect the reported verdicts.
Our results thus characterize call-graph-guided evidence acquisition rather than the full range of analyses a CPG can support.

Alongside the call-graph projection, the model receives PE metadata comprising the file and optional headers, the section table and per-section entropies, the import and export tables, embedded resources, and file hashes.
\Cref{sec:evidence} evaluates the effect of including this metadata by comparing the fixed pipeline with and without it.

\subsection{Triage as Budgeted Evidence Acquisition}

Because the system cannot inspect every recovered function, it must decide where to spend a limited number of function analyses.
We therefore formulate triage as evidence acquisition under a budget: the pipeline selects a subset of functions, records the resulting analyses in a shared dossier, and derives its verdict only from the evidence acquired within that allowance.

The budget is divided across the three workflow stages:
\begin{itemize}
    \item The \emph{seeding budget} determines how many functions are selected by the fixed ranking heuristic and analyzed to initialize the evidence dossier.
    \item The \emph{exploration budget} limits how many additional functions the agent may analyze while following leads through the call graph.
    \item The \emph{reporting budget} limits how many read-only dossier inspections the reporting agent may perform before returning its verdict.
\end{itemize}

The exploration and reporting stages may stop before exhausting their respective allowances.
This design holds the maximum evidence budget and available actions fixed across models while allowing their stopping behavior and choice of exploration targets to differ.

\subsubsection{Seed Analysis}\label{sec:eps}
Seed analysis initializes the evidence dossier.
For a given binary, every evaluated model starts from the same ranked set of seed functions, so initial selection cannot confound the cross-model comparison, though the model-generated analyses of those functions may differ.

We use a recall-oriented seed policy.
In the development experiments, later exploration rarely recovered malware whose relevant evidence was missed during seeding, as \Cref{sec:eval-exploration} shows.
The selected policy therefore favors an initial operating point with high malware recall, leaving later stages to resolve false alarms when additional context permits.

Each function receives an Evidence Priority Score (EPS), defined as a weighted combination of features derived from its decompiled body and its position in the call graph.
\begin{equation}
\mathrm{EPS}(f) = 0.25\,\mathrm{PR}_{\text{out}}(f) + 0.50\,B(f) + 0.25\,C(f).
\end{equation}

$\mathrm{PR}_{\text{out}}$ is PageRank computed on the reverse call graph and measures the structural prominence of $f$ as a caller.
The boundary indicator $B(f)$ equals one when $f$ calls at least one imported API and has either zero internal in-degree or zero internal out-degree.
The degree condition constrains only $f$'s position in the internal call graph, so a root that no recovered function calls qualifies even when it calls internal functions alongside imported APIs.

$C(f)$ is the log-scaled length of the decompiled function body, normalized within the binary:
\[
C(f) = \frac{\log(1+|\mathrm{code}(f)|)}
            {\max_{g \in F} \log(1+|\mathrm{code}(g)|)},
\]
where $F$ is the set of recovered functions in the binary. 
It is a simple proxy for the amount of code available for analysis.

EPS thus combines three inexpensive signals: caller prominence, interaction with imported APIs at a call-graph boundary, and decompiled body length.
The weights were selected from a grid search over 70 weight vectors on the training split, reported in Appendix~\ref{app:eps-grid}. 
We do not treat this weighting as uniquely optimal, and compare EPS with several alternative selectors in \Cref{sec:funcprior}.

Each of the $k$ highest-scoring seed functions is analyzed once by the evaluated LLM. 
The model receives the decompiled function body and is asked to separate the mechanism implemented by the code from any inference about its intent.
Because many low-level operations are dual-use, the prompt asks the model to consider plausible benign explanations rather than treating a suspicious mechanism as sufficient evidence of malware. 
For each function, the model returns a summary, a suspicion score between zero and one, the capabilities supported by the code, and the strongest benign explanations.
These fields form the function's record in the evidence dossier.

\subsubsection{Evidence Exploration}\label{sec:exploration}

During exploration, an agent may expand the initial dossier by acquiring evidence from functions that were not selected as seeds.
Its state consists of a running scratchpad containing its current hypothesis, a history of previous actions, and a summary of each analyzed function.
At every step, the agent also receives the latest tool result, the PE metadata, and its remaining function-analysis budget.

The agent may \texttt{analyze} an unread function, list a function's callers or callees, find functions that call a specified API, or stop.
Only \texttt{analyze} consumes the function-analysis budget, while graph queries are free.
The agent is instructed to explore when the existing dossier is insufficient and may stop early when it considers the evidence sufficient.
Unused budget receives no explicit reward, so differences in budget use reflect how models respond to the same evidence and stopping instruction.
To bound query-only exploration, the stage terminates after ten consecutive actions without a successful \texttt{analyze} call.
It otherwise ends when the agent stops or exhausts its analysis budget.

\subsubsection{Verdict Reporting}\label{sec:verdict}
The reporting agent operates on a frozen evidence dossier. 
It cannot analyze additional code or issue new graph queries and may only inspect the evidence collected during seeding and exploration.
This separation prevents the reporting stage from changing the evidence set while adjudicating it.

The agent must return one of three verdicts.
\emph{Benign} and \emph{malicious} are committed decisions. 
\emph{Suspicious} indicates that the collected evidence does not support either committed verdict and is interpreted as escalation to deeper analysis.
The prompt cautions the model that an unsuccessful graph query is not evidence that a capability is absent and that high confidence in a benign verdict requires inspection of the likely execution path.
The final structured report contains the verdict and confidence, the supporting functions and evidence, the identified capabilities, a concise rationale, and a recommended next step.

All evaluated models receive the same zero-shot stage instructions, action schemas, and budget limits.
No prompt or tool interface is tuned for an individual model.
The complete prompts and structured output schemas are included in the accompanying artifact.

\section{Experimental Setup}
This section describes how we evaluate TRIAGE-CPG: \Cref{sec:dataset} introduces the dataset and its splits, \Cref{sec:models} the models and locked protocol, and \Cref{sec:metrics} the metrics.

\subsection{Dataset and Splits}\label{sec:dataset}
Our dataset consists of \num{1000} Windows PE binaries sampled from VirusTotal.
When constructing the dataset, we required zero VirusTotal detections for benign and at least ten for malware.
We use these high-consensus criteria as evaluation labels rather than definitive ground truth.
Samples with 1--9 detections were excluded, so our results do not characterize binaries on which VirusTotal engines disagree. 

We include only binaries that yield a CPG with at least one valid function and exclude .NET binaries, which the CPG implementation does not support.
The evaluation is therefore conditional on successful CPG extraction.
The samples are balanced by label (500 benign, 500 malicious) and across five file-size quantiles computed separately for each class.
This ensures that neither class concentrates in a single size regime.

We partition the data into three disjoint splits: a \emph{train} split (200) for selecting the EPS weights, a \emph{development} split (200) for ablations and budget selection, and a locked \emph{test} split (600) used once for the cross-model comparison.
This gives 20 samples per label--quantile cell in train and development splits, and 60 in the test split.

\begin{figure}[!ht]
\centering
\includegraphics[width=\columnwidth]{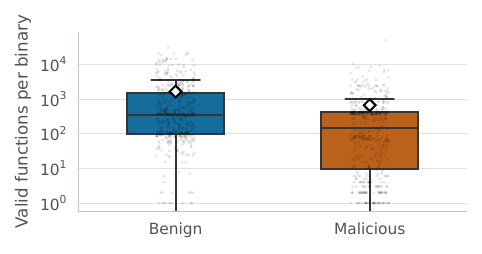}
\Description{Two box-and-jitter plots of valid functions per binary on a logarithmic vertical axis, from one to more than ten thousand. 
The benign distribution, shown in blue, lies higher and is more spread out than the malicious distribution, shown in orange. 
The benign median is 364 recovered functions and the malicious median is 207.
Both distributions are strongly right-skewed, with individual binaries reaching 49,519 functions.}
\caption{
Recovered functions per binary by label (log scale). Boxes show the interquartile range and median, diamonds the mean, and dots individual binaries. Counts reach \num{49519} (median 304; mean \num{1208}), with benign binaries denser than malicious ones.
}
\label{fig:function-count-distribution}
\end{figure}

Overall, the binaries have a median of about 300 recovered functions, a mean near \num{1200}, and a long tail reaching roughly \num{49500}.
File sizes span roughly 1\,KB to 141\,MB (median 501\,KB), and benign binaries are larger than malicious ones (median 682 vs.\ 354\,KB).
\Cref{fig:function-count-distribution} shows the function counts by label: both distributions are heavily right-skewed, and benign binaries are also function-denser than malicious ones (median 364 vs.\ 207).
This scale motivates budgeted inspection, since whole-binary prompting or exhaustive per-function analysis would require processing hundreds to thousands of functions per sample.

\subsection{Models and Protocol}\label{sec:models}
We selected eight models based on their availability on Vertex AI at experiment time.
These comprise three closed-source Gemini models (Gemini 3.1 Pro, Gemini 3.5 Flash, and Gemini 3.1 Flash-Lite) and five open-weight models spanning several families and sizes (DeepSeek V4 Pro, GLM 5.2, Gemma 4 31B, Qwen 3.6 27B, and Gemma 4 26B).
The selection is not exhaustive and reflects platform availability rather than a global model ranking.

All models ran under the same configuration, selected on the development split and locked before the test runs.
The seed stage selected $k = \min(\max(\lceil 0.1|F|\rceil, 10), 75, |F|)$ functions by EPS, where $|F|$ is the number of recovered functions.
The exploration agent received an allowance of 30 \texttt{analyze} calls and was shown the remaining allowance at every step.
Exploration ended when the agent selected \texttt{done}, exhausted this allowance, or performed ten consecutive actions without a successful \texttt{analyze} call.
The reporting agent received an allowance of 20 dossier-inspection actions.
PE metadata was included by default and ablated separately in \Cref{tab:evidence-inputs}.
These values define the common operating point used for comparison, not a globally optimal configuration.

Each model was evaluated once per binary at temperature 0, without model-specific tuning of prompts, budgets, or decoding parameters.
Because the pipeline makes dozens of model calls per binary, repeating the complete test evaluation would require tens of thousands of additional calls per model.
Even at temperature 0, API-served models do not produce deterministic outputs.
We therefore estimated run-to-run variation by repeating the fixed Gemini 3.1 Flash-Lite configuration three times on the development split.
The runs agreed on 91\% of verdicts and spanned 1.5 balanced-accuracy points, with most disagreements involving a change to or from the \emph{suspicious} verdict.
We use this 1.5-point span as an empirical reference and avoid drawing conclusions from smaller cross-model differences.

\subsection{Metrics}\label{sec:metrics}
We evaluate the ternary verdicts under two readings.
Under the \emph{strict} reading, only a committed \emph{malicious} verdict counts as a positive prediction. 
This reading isolates committed malware calls from cases where the model defers as \emph{suspicious}.
Under the \emph{triage} reading, a malware sample is handled correctly if it is deemed malicious or escalated, while a benign sample is handled correctly only when it is cleared.
We call balanced accuracy under this reading the \emph{triage balanced accuracy} (tBA) and use it as the primary triage-oriented metric, since it captures silent malware clears and benign samples routed to further analysis.
Writing $v$ for the verdict,
\[
\mathrm{tBA} = \tfrac{1}{2}\left(\Pr[v \neq \textit{benign} \mid \text{malware}] + \Pr[v = \textit{benign} \mid \text{benign}]\right),
\]
which is equivalently the balanced accuracy of the clearing decision: a malware sample must not be cleared, while a benign sample must be.
A model that escalates every sample obtains a tBA of 0.5, providing no improvement over the chance-level balanced-accuracy baseline.
By design, tBA does not distinguish a committed malicious verdict from an escalation on a malware sample, so we never report it alone: the strict reading identifies committed malicious decisions, while the escalation quantities below characterize the queue created by suspicious verdicts.
The two readings are the extreme points of a cost-charged family in the error--reject tradition \cite{chowOptimumRecognitionError1970a}, so any intermediate assumption about the cost of an escalation yields a value between the \emph{strict} and \emph{triage} numbers we report. The gap between them measures how much a model's performance rests on escalation rather than committed verdicts.
From the confusion matrix of either reading, we report recall (the fraction of malware caught), specificity (the fraction of benign samples cleared), precision, and balanced accuracy. 
Since the dataset is balanced, accuracy and balanced accuracy coincide, and we report balanced accuracy throughout.

For the suspicious verdict itself, we report two escalation quantities.
\emph{Coverage} is the share of samples marked suspicious.
\emph{Precision} is the fraction of suspicious samples that are malware. Because the benchmark is balanced at a 0.5 malware base rate, a precision above 50\% means the escalation queue is richer in malware than the dataset as a whole.

\section{Evaluation}\label{sec:evaluation}
We organize the evaluation around four research questions, each answered in a dedicated subsection:
\begin{enumerate}[label=\textbf{RQ\arabic*},ref=RQ\arabic*]
    \item\label{rqEvidence} What is the impact of the PE metadata on the verdict?
    \item\label{rqEPS} Which functions should the seed budget be spent on first?
    \item\label{rqExploration} What does the exploration stage add? 
    \item\label{rqModels} How do the models differ in their triage behavior under the same frozen protocol?
\end{enumerate}
RQ1--RQ3 characterize design choices in the fixed workflow using the development split and Gemini 3.1 Flash-Lite, which we chose for its low cost and fast inference. Their outcome is the frozen configuration of \Cref{sec:models}.
These design choices were made under the strict reading and before any test-split run, so the primary tBA metric played no role in selecting the configuration.

RQ4 evaluates that frozen configuration across eight models on the locked test split, which is touched exactly once.
The escalation and cascade analyses are post-hoc analyses of the locked test outputs, not newly tuned decision rules.

\subsection{Impact of PE Metadata}\label{sec:evidence}
Before asking which functions to inspect, we check what evidence the model needs to reach a verdict.
\Cref{tab:evidence-inputs} compares the triage performance when using only the PE metadata, the CPG alone, and the CPG with the PE metadata.

\begin{table}[!ht]
\centering
\setlength{\tabcolsep}{4.5pt}
\caption{Development-split evidence ablation (Gemini 3.1 Flash-Lite). Metadata-only rows use both verdict readings; code rows use fixed seeding without exploration under the strict reading.}
\label{tab:evidence-inputs}
\begin{tabular}{lrrrr}
\toprule
Input & Bal. Acc. & Recall & Spec. & FP/FN \\
\midrule
Metadata only (strict) & 0.505 & 0.010 & 1.000 & 0/99 \\
Metadata only (triage) & 0.770 & 1.000 & 0.540 & 46/0 \\
\midrule
Code only              & 0.670 & 0.900 & 0.440 & 56/10 \\
Code + metadata        & 0.795 & 0.890 & 0.700 & 30/11 \\
\bottomrule
\end{tabular}
\end{table}
Given only the PE metadata, the model commits to a malicious verdict for one of the 100 malware samples (strict recall 0.010).
It marks 145 of the 200 binaries (72.5\%) as suspicious, 99 of the malware samples and 46 benign ones, which drives the triage recall to 1.000 at a specificity of only 0.540.
In this development configuration, metadata alone raises suspicion broadly but rarely supports a committed malicious verdict.
Code evidence is more discriminative, while adding PE metadata reduces false positives from 56 to 30 at nearly unchanged recall.
Thus, PE metadata acts as a benign prior in this experiment, while code supplies most of the verdict-separating evidence.

\begin{framed}
\noindent\textbf{Answer to \ref{rqEvidence}:}
Code supplies most of the verdict-separating evidence, while PE metadata primarily reduces false positives, so we retain both in the fixed protocol.
\end{framed}

\subsection{EPS as a Prioritization Signal}\label{sec:funcprior}
The seed stage must choose which functions to analyze before any model call has been made.
For each binary in the development split, we rank the recovered functions with EPS and with six alternatives: largest code body, out-degree, in-degree, number of API calls, EPS without its code term, and a random selector, run five times with different draws.
Each selector's top-ranked functions are analyzed and adjudicated directly, with the exploration stage disabled, so the comparison isolates seeding quality.\footnote{The selection experiments use the proportional seed rule with a floor of 10 and a cap of 100 seeds. The fixed protocol caps at 75, an operating-point choice examined below.}

\begin{figure}[t]
\centering
\includegraphics[width=\columnwidth]{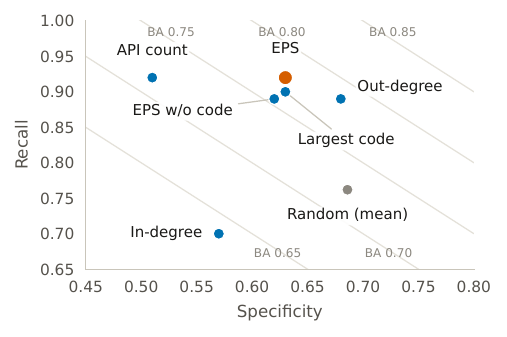}
\Description{Scatter plot of recall against specificity for seven seed-function selectors, overlaid with diagonal reference lines of constant balanced accuracy from 0.65 to 0.85.
EPS has recall 0.92 at specificity 0.63, and largest code and EPS without the code term sit just below it at recall 0.89 to 0.90 and specificity 0.62 to 0.63.
Out-degree has recall 0.89 at the higher specificity of 0.68.
These four selectors all fall between the 0.75 and 0.80 reference lines.
API count has the same recall as EPS, 0.92, but a much lower specificity of 0.51, placing it near the 0.70 line.
The random mean lies at recall 0.76 and specificity 0.69, also near the 0.70 line, and in-degree is lowest at recall 0.70 and specificity 0.57.}
\caption{
Development-split seed-selector performance (strict verdict, exploration off). Diagonals mark constant balanced accuracy, so selectors on a common line perform equally well. The random baseline is the mean of five runs.
}
\label{fig:function-selection}
\end{figure}

\Cref{fig:function-selection} plots each selector's recall and specificity, and the comparison contains a null result that we consider informative: the tested inexpensive ranking signals perform similarly on the development split.
EPS (0.775 balanced accuracy), out-degree (0.785), and largest code body (0.765) land within two points of each other, and even the mean random selector reaches 0.724.
The competitive performance of the largest-code selector is consistent with code and graph evidence being concentrated in larger functions: the largest tenth of a binary's functions holds a median of 26\% of its external API calls and 26\% of its internal call edges, a 2.6--2.9$\times$ enrichment over proportional coverage.
The selectors also differ in their position on the recall--specificity frontier, as depicted in \Cref{fig:function-selection}.
EPS is the most recall-oriented selector,  catching 92\% of malware at 0.63 specificity under the strict reading, while the only other alternative matching this recall, ranking by API count, results in a specificity of 0.51.
This position is not an artifact of picking the same functions as the other heuristics. Per binary, only a median 33\% of EPS seeds appear in the largest-code seed set and 23\% in the out-degree seed set.
We argue in \Cref{sec:eval-exploration} that a recall-heavy seeder is the right choice for this pipeline, because the exploration stage recovers specificity but does not recover missed malware.

\paragraph{Seed-budget sensitivity.}
The fixed schedule analyzes 10\% of the recovered functions, with a floor of 10 and a cap of 75.
The fixed-count sweep across binary-size classes in Appendix~\ref{app:func-select} supports the proportional term: no single seed count serves all sizes, and performance decreases when too many low-ranked functions are included.
The choice of upper cap is less clear-cut.
\Cref{tab:seed-cap-sensitivity} varies it under the full pipeline on the development split.
Lowering the cap to 50 improves balanced accuracy by cutting false positives, at the price of two more missed malware samples under the strict reading.
Raising it to 100 recovers additional malware but adds false positives and cost, with balanced accuracy unchanged from the 75 limit, and the change helps and hurts nine binaries each.
No fixed limit dominates all goals.
We freeze cap 75 for the locked evaluation as a recall-biased, cost-controlled middle point, consistent with the recall-first result of \Cref{sec:eval-exploration}.

\begin{table}[!ht]
\centering
\setlength{\tabcolsep}{3.5pt}
\caption{Development-split seed-cap sensitivity under the full pipeline and strict verdict; all other configuration values are fixed.}
\label{tab:seed-cap-sensitivity}
\begin{tabular}{lrrrrr}
\toprule
Cap & Avg.\ analyzed & Bal.\ Acc. & Recall & Spec. & FP/FN \\
\midrule
50  & 35.5 & 0.820 & 0.870 & 0.770 & 23/13 \\
75  & 44.8 & 0.790 & 0.890 & 0.690 & 31/11 \\
100 & 52.0 & 0.790 & 0.910 & 0.670 & 33/9 \\
\bottomrule
\end{tabular}
\end{table}

\begin{framed}
\noindent\textbf{Answer to \ref{rqEPS}:}
Several inexpensive static selectors perform similarly. 
We select EPS because it occupies the recall-oriented end of their recall--specificity frontier, which is preferable when later exploration rarely recovers missed malware.
\end{framed}

\subsection{Impact of the Exploration Stage}\label{sec:eval-exploration}
The exploration stage lets the agent read functions the static ranking did not select.
We measure its contribution on the development split in three ways: switching it off, replacing the agent's target choice with random draws, and re-running the seed selectors of \Cref{sec:funcprior} under the full pipeline.

\paragraph{Exploration primarily corrects false positives.}
\Cref{tab:exploration-onoff} pairs runs with and without exploration across nine seed and reporting configurations, and counts per-binary verdict flips.
The effect appears to concentrate on the benign side.
Pooled over the nine configurations and counting each binary once by its net flip direction, exploration corrects 30 benign binaries and hurts 11 (sign test clustered by binary, $p=0.004$), while for malware it shows no consistent effect (corrects 9, hurts 18, $p=0.12$).
The fixed configuration shows the same picture, clearing ten additional benign binaries while catching no additional malware. 
In these configurations, the observable effect of exploration is therefore primarily the resolution of false alarms rather than the recovery of malware missed during seeding.

\paragraph{Guided targeting does not improve accuracy but stops early.}
A natural question is whether the agent's choice of exploration targets matters at all.
We replace the agent's target choice with uniformly random draws over unanalyzed functions in three full-budget and three action-matched random arms (\Cref{tab:exploration-policy}).
On accuracy, the comparison is mixed, and no difference reaches significance.
This null result bounds the contribution of agentic target selection in this setting.
The agent's remaining advantage is economy: it decides when to stop and halts after an average of 5.2 additional function reads per binary, whereas the random policies have no stopping rule and spend their full allowance of 11.3 to 25.4 reads without an accuracy gain.
We return to what the agent actually read, across all eight models, in \Cref{sec:final}.

\paragraph{Exploration rarely recovers seed misses.}
Re-running the seed selectors under the full pipeline, reported in \Cref{tab:exploration-selectors}, connects the two findings above.
Exploration converges the specificity of the three selectors into a narrow band (0.70--0.72), but the recall differences persist: EPS misses 11 of 100 malware samples, the largest-code selector misses 18, and the random selector misses 13.\footnote{The full-pipeline random arm reuses the strongest of the five random runs of \Cref{sec:funcprior} (seed-only balanced accuracy 0.760 of a 0.650--0.760 spread), so it is an optimistic estimate of random seeding.}
On the binaries where EPS and largest-code disagree, largest-code misses eight malware samples that EPS catches, against one in the other direction.
This asymmetry is the central lesson of the design experiments.
A false positive in the dossier can be resolved downstream, but a binary whose seeds show nothing suspicious gives the agent no lead to follow, and the reporting stage cannot recover what was never read.
Under this asymmetry, the seed ranker should maximize recall at acceptable specificity, which is exactly the operating point EPS occupies.

\begin{framed}
\noindent\textbf{Answer to \ref{rqExploration}:}
Exploration primarily improves specificity rather than recall: it resolves some false alarms and often uses little of its allowance, but does not consistently recover malware missed during seeding. This supports the recall-oriented seeding choice of RQ2.    
\end{framed}

\subsection{Comparison of Models}
\label{sec:final}
\begin{figure*}[t]
\centering
\includegraphics[width=\textwidth]{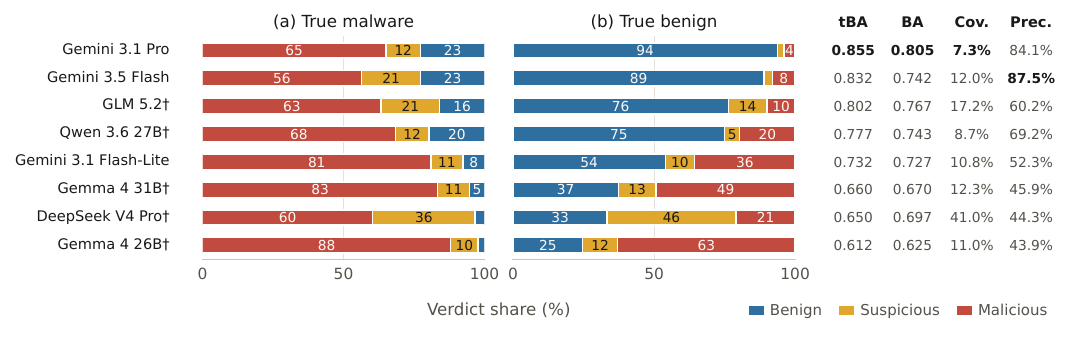}
\Description{Two panels of horizontal stacked bars, one row per model, ordered by triage balanced accuracy from Gemini 3.1 Pro at the top to Gemma 4 26B at the bottom.
Panel (a) covers the 300 true-malware binaries and stacks the malicious, suspicious, and benign shares from left to right.
Panel (b) covers the 300 true-benign binaries and stacks the benign, suspicious, and malicious shares from left to right.
In both panels the leftmost segment is the correct verdict, so bars that are long on the left indicate better performance.
Gemini 3.1 Pro has a short correct segment in panel (a) but the longest in panel (b), while Gemma 4 26B shows the opposite pattern.
DeepSeek V4 Pro has by far the widest suspicious segment in both panels.}
\caption{Verdict composition on the locked test set, one row per model sorted by tBA, over (a) the 300 true-malware and (b) the 300 true-benign binaries. Both panels stack the correct verdict first (\emph{malicious} leftmost in (a), \emph{benign} in (b)). Each reading is a cut point: strict recall is the malicious segment of (a), and triage recall adds the adjacent suspicious one. The right block reports triage and strict balanced accuracy (tBA, BA) and the coverage and precision of the suspicious queue. Open-weight models are marked \textsuperscript{\dag}. Short names: Pro/Flash/FL = Gemini 3.1 Pro/3.5 Flash/3.1 Flash-Lite, DSv4 = DeepSeek V4 Pro, G31/G26 = Gemma 4 31B/26B.}
\label{fig:model-policy}
\end{figure*}

With the configuration frozen, we ran all eight models once over the 600 locked test binaries.
\Cref{fig:model-policy} shows how each model distributed its three verdicts over the true-malware and true-benign halves of the split, together with the summary metrics derived from those distributions.

\paragraph{Triage operating modes.}
The models do not converge on a shared behavior and do not split into closed-source and open-weight groups.
At the high-specificity end, Gemini 3.1 Pro commits a malicious verdict for 65\% of malware while clearing 93.7\% of benign samples, the highest strict balanced accuracy (0.805) and tBA (0.855) under the fixed protocol.
Gemini 3.5 Flash, GLM 5.2, and Qwen 3.6 27B also occupy comparatively high-specificity operating points.
At the high-recall end, Gemma 4 26B commits to a malicious verdict for 88\% of malware but reaches only 0.370 specificity, with Gemma 4 31B and Gemini 3.1 Flash-Lite showing similar recall-oriented behavior.\footnote{Gemini 3.1 Flash-Lite was used to select the fixed configuration on the development split, where it reached 0.800 balanced accuracy. The cross-model comparison uses its locked test result of 0.727.}

Neither model provenance nor nominal size explains these operating points.
The open-weight GLM 5.2 and closed-source Gemini 3.1 Pro produce the same ternary verdict on 78.3\% of test binaries, while similarly sized open-weight models occupy different parts of the recall--specificity spectrum.
The highest-scoring open-weight model, GLM 5.2, lies 5.3 tBA points below Gemini 3.1 Pro.
The cross-model differences are also substantially larger than the run-to-run variation observed in \Cref{sec:models}: repeated runs of one model agree on 91\% of the development-split verdicts, while Gemini 3.5 Flash and Gemma 4 26B agree on just 48.2\% of the test set.

\begin{figure}[t]
\centering
\includegraphics[width=\columnwidth]{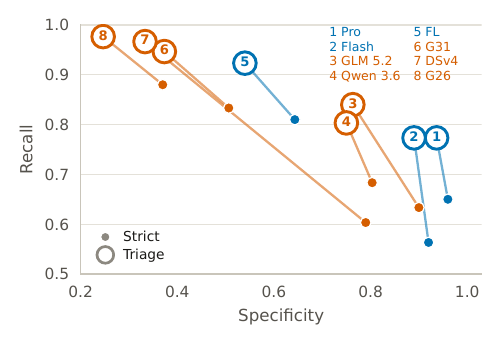}
\Description{Scatter plot of recall against specificity for the eight models. Each model is drawn as a line segment from its strict operating point, a small filled dot, to its triage operating point, a numbered open circle, colored blue for closed-source and orange for open-weight models. A key numbers the models one to eight by triage balanced accuracy. Reinterpreting suspicious verdicts as escalation moves every model up and to the left, raising recall and lowering specificity, by amounts that vary widely across models.}
\caption{Strict and triage operating points on the locked test set. Each model is a segment from its strict verdict (small filled dot) to its triage reading (numbered open circle). Models are numbered 1--8 by tBA as in \Cref{fig:model-policy}, closed-source blue and open-weight orange.}
\label{fig:operating-points}
\end{figure}

\paragraph{Suspicious verdict as escalation policy.}
\Cref{fig:operating-points} plots each model's strict and triage operating points, and comparing the two shows how reinterpreting suspicious verdicts changes recall and specificity.
For Gemini 3.1 Pro, including suspicious verdicts raises recall from 0.650 to 0.773 while reducing specificity from 0.960 to 0.937.
For DeepSeek V4 Pro, recall rises from 0.603 to 0.967, but specificity falls from 0.790 to 0.333.

The suspicious queues also differ substantially in composition. 
Gemini 3.5 Flash and Gemini 3.1 Pro escalate 12.0\% and 7.3\% of samples at 87.5\% and 84.1\% precision, respectively. 
In contrast, DeepSeek V4 Pro escalates 41.0\% of samples at 44.3\% precision.
The suspicious verdict therefore cannot be interpreted uniformly across models: its frequency and malware enrichment depend strongly on the selected model.

\paragraph{Post-hoc suspicious queue pruning.}
We examine, post hoc, whether a second model can prune an existing suspicious queue, as depicted in \Cref{tab:cascade-top}.
Routing Gemini 3.1 Pro's queue through Gemini 3.5 Flash, the highest-tBA pair among the 56 ordered pairs, shrinks the queue from 44 to 35 samples and raises its precision from 0.841 to 0.943, at the cost of two more missed malware samples.
Among open-weight pairs, routing GLM 5.2's queue through Gemma 4 26B reduces it from 103 to 25 samples and raises precision from 0.602 to 0.840 while missing one additional malware.
The overall-best cascade raises tBA by only 0.005, within the observed run-to-run variation.
Cascading therefore prunes analyst load at a pair-dependent cost in silent misses rather than improving the classifier.

\begin{table}[h!]
\centering
\small
\setlength{\tabcolsep}{5pt}
\caption{Post-hoc cascades on locked test outputs: top overall and open-weight pairs by tBA, with single-model baselines. Lost is malware cleared by the second model; abbreviations follow \Cref{fig:model-policy}.}
\label{tab:cascade-top}
\begin{tabular}{rlrrrrr}
\toprule
\# & System & BA & tBA & Queue & Prec. & Lost \\
\midrule
   & \textit{Pro alone}   & .805 & .855 &  44 & .841 & --  \\
 1 & Pro$\to$Flash        & .808 & .860 &  35 & .943 & 2   \\
 2 & Pro$\to$GLM 5.2      & .815 & .857 &  29 & .931 & 2   \\
 3 & Pro$\to$G31          & .832 & .855 &  18 & .889 & 1   \\
 4 & Pro$\to$G26          & .828 & .855 &  22 & .864 & 0   \\
 5 & Pro$\to$DSv4         & .808 & .853 &  35 & .886 & 2   \\
\midrule
   & \textit{GLM 5.2 alone} & .767 & .802 & 103 & .602 & -- \\
18 & GLM 5.2$\to$DSv4     & .777 & .807 &  80 & .613 & 2   \\
19 & GLM 5.2$\to$G26      & .775 & .803 &  25 & .840 & 1   \\
20 & GLM 5.2$\to$Qwen 3.6 & .782 & .803 &  25 & .760 & 22  \\
21 & GLM 5.2$\to$G31      & .773 & .802 &  31 & .774 & 4   \\
24 & Qwen 3.6$\to$GLM 5.2 & .767 & .788 &  25 & .760 & 1   \\
\bottomrule
\end{tabular}
\end{table}

\paragraph{How the budget is spent.}
The models also differ in how they use the exploration stage, which \Cref{tab:exploration-audit} breaks down per model.
GLM 5.2, Qwen 3.6 27B, and Gemini 3.1 Pro analyze only 3.6 to 5.4 new functions per sample and stop on their own in 77--90\% of runs, while Gemini 3.5 Flash and Gemma 4 26B analyze 9.2 and 14.0 and mostly run their budget out.
Target selection styles differ as well: Gemini 3.1 Pro and Qwen 3.6 27B derive about 55\% of their targets from their own scratchpad reasoning, Gemma 4 26B follows callees more often (31\%), and Gemma 4 31B and DeepSeek V4 Pro lean on prior analysis outputs (about 29\%).
Despite these styles, all eight models leave the static ranking behind: 35--51\% of newly analyzed functions rank outside the top 150 by EPS, and 42--54\% are not adjacent to any initial seed.
Exploration evidence is frequently used in the final reports.
Functions read during exploration are cited in the final report 3.7--7.7$\times$ more often than seed functions, as shown in \Cref{tab:exploration-audit}.
This citation frequency measures whether the models use the evidence, not whether the cited conclusions are correct.
Reading more is not itself an advantage: the highest-tBA model is among the most frugal and the lowest-tBA model among the most prolific, so there is no monotonic relationship between exploration reads and triage performance under this fixed protocol.

\begin{table}[!ht]
\centering
\small
\setlength{\tabcolsep}{2pt}
\caption{Exploration and reporting behavior. Reads = mean exploration analyses; Stop = explicit \texttt{done}; Scratch./Callees = target-source shares; $>$150 = outside the EPS top 150; Non-adj. = not adjacent to a seed; Rep. = mean reporting inspections; Cited = explored-to-seed citation-rate ratio. Abbreviations follow \Cref{fig:model-policy}.}
\label{tab:exploration-audit}
\begin{tabular}{lrrrrrrrr}
\toprule
Model & Reads & Stop & Scratch. & Callees & $>$150 & Non-adj. & Rep. & Cited \\
\midrule
Pro      &  5.4 & 87\% & 56\% & 21\% & 47\% & 46\% & 1.8 & 7.7 \\
Flash    &  9.2 & 45\% & 39\% & 22\% & 43\% & 54\% & 5.9 & 4.1 \\
FL       &  6.2 & 86\% & 33\% & 25\% & 43\% & 45\% & 0.0 & 7.2 \\
DSv4     &  7.1 & 31\% & 45\% & 14\% & 35\% & 46\% & 3.6 & 4.1 \\
GLM 5.2  &  3.6 & 90\% & 45\% & 11\% & 43\% & 42\% & 1.4 & 5.6 \\
G31      &  8.1 & 75\% & 30\% & 21\% & 49\% & 47\% & 0.3 & 4.8 \\
Qwen 3.6 &  4.5 & 77\% & 55\% & 11\% & 43\% & 46\% & 0.4 & 6.5 \\
G26      & 14.0 & 34\% & 33\% & 31\% & 51\% & 48\% & 2.7 & 3.7 \\
\bottomrule
\end{tabular}
\end{table}

The reporting budget, in contrast, is used sparingly.
Six of the eight models take a median of at most one inspection action before committing to a verdict, and no model exhausts the 20-action budget in even 4\% of its runs, so the final stage adjudicates the dossier essentially as the earlier stages handed it over.

\begin{framed}
\noindent\textbf{Answer to \ref{rqModels}:}
Under identical evidence and action budgets, the models occupy different recall--specificity and escalation operating points. No model dominates across malware recall, benign clearance, and escalation-queue size, so model selection is itself part of the triage policy.  
\end{framed}

\section{Discussion and Limitations}
Because the models implement different policies under identical evidence, the comparison in \Cref{sec:final} maps operating points under one fixed protocol rather than ranking model quality or fixing a deployment configuration. Per-model tuning could move the individual points.

\subsection{Quality of the Reports}
\label{sec:report}
In addition to the verdict, TRIAGE-CPG produces a structured behavioral report.
We assess these reports with an automated proxy judge (Gemini 3.1 Flash-Lite) that compares each against the threat-intelligence evidence derived from VirusTotal for the same sample, returning yes, no, or unknown agreement on the reported malware type, family, capabilities, and role, and flagging over- and under-claiming.\footnote{The judge is one of the eight evaluated models. This assessment measures agreement with the VirusTotal reference evidence rather than independent expert validation.}

\begin{figure}[t]
\centering
\includegraphics[width=\columnwidth]{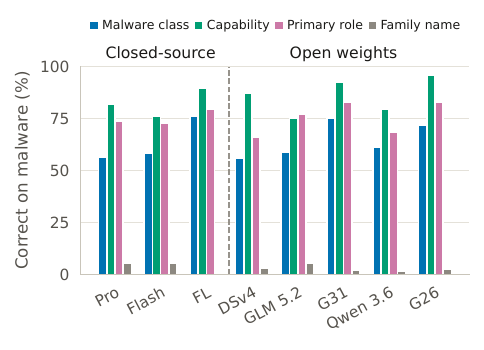}
\Description{Grouped bar chart of agreement percentages on malware samples, one group of four bars per model, with the three closed-source models on the left and the five open-weight models on the right, separated by a dashed divider.
Within each group the bars are malware class, capability, primary role, and family name.
Capability agreement is highest for every model, from roughly 75 to 96 percent.
Primary-role agreement is around 66 to 83 percent, followed by malware-class agreement, around 56 to 76 percent.
Family-name agreement is near zero for every model, roughly 1 to 6 percent, leaving a wide gap below the other three bars.}
\caption{
Agreement between report characterizations and external VirusTotal-derived threat-intelligence evidence on the malware half of the locked test set, one group of bars per model, split into closed-source and open-weight groups.
}
\label{fig:report-characterization}
\end{figure}

\Cref{fig:report-characterization} shows the result on the malware half of the test set. 
Reported capabilities agree with the external consensus in 75--96\% of the cases, the role in 66--83\%, and the malware type in 56--76\%.
Family attribution agrees in only 1--6\% of cases, reflecting its dependence on vendor naming and campaign context the static code evidence does not provide.
On malware, reports under-claim more often than they over-claim (45--56\% vs.\ 13--22\%).
On benign samples, over-claiming averages 35\% across models and is highest for models that also issue malicious verdicts broadly.

These proxy results indicate that capability- and role-level descriptions agree with the external evidence more consistently than family names, but not that the reports improve analyst decisions or reduce analysis time, which would require a human study.

\subsection{Limitations of Static Triage}
\label{sec:failure}
The test-set errors reveal two recurring limitations of the static CPG-based workflow: large binaries and code that hides its relevant behavior behind a wrapper.

\paragraph{Failure modes by file size.}
Performance becomes less balanced as binary size increases, but the direction of the errors follows each model's operating point.
Under the strict reading, Gemini 3.1 Pro maintains specificity of at least 0.95 across all size quantiles, while its malware recall falls from 0.783 in the smallest quantile to 0.33 in the largest. 
GLM 5.2 shows a similar pattern: recall falls from 0.733 to 0.350 while specificity remains at least 0.80.
The recall-oriented Gemma 4 26B shows the opposite pattern, maintaining recall between 0.80 and 0.95 while its specificity falls from 0.733 to 0.167.
Larger binaries therefore amplify the model-specific recall--specificity trade-offs observed in \Cref{sec:final}.

\paragraph{Packers, protectors, and installers.}
Static triage is inherently limited on samples that hide their payload behind a delivery shell such as a packer, protector, or installer, because malicious and benign software then rely on the same mechanisms.
We approximate this class with VirusTotal's Detect-It-Easy results \cite{horsHorsicqDetectItEasy2026}, marking such binaries as wrappers.
Droppers and loaders have no equally consistent marker in our data, so the resulting wrapper rate is a lower bound.
\Cref{tab:failure-semantics} reports the wrapper share within each verdict cell.

Wrapped binaries constitute 31\% of the test set and occur at similar rates among benign and malicious samples (33\% against 29\%).
They are less common among the correctly flagged malware and cleared benign samples (19\% and 28\%) but more common among missed malware and false positives (44\% each). 
They are particularly concentrated among escalated malware, of which 61\% is wrapped.
These results identify wrappers as a shared difficult class for both verdict errors and escalation.

Under the intended triage interpretation, suspicious verdicts on wrapped samples would be routed to unpacking or dynamic analysis.
A future unpack-and-re-extract action that constructs a new CPG from the recovered payload could directly address this limitation.

\begin{table}[!ht]
\centering
\setlength{\tabcolsep}{5pt}
\small
\caption{Wrapper share within each verdict cell. Detect-It-Easy identifies packers, protectors, or installers; the baseline is 29\% for malware and 33\% for benign samples. Per-cell counts vary, so the mean row is the robust reference.}
\label{tab:failure-semantics}
\begin{tabular}{lrrrrrr}
\toprule
& \multicolumn{3}{c}{Malware} & \multicolumn{3}{c}{Benign} \\
\cmidrule(lr){2-4}\cmidrule(lr){5-7}
Model & TP & Susp. & FN & TN & Susp. & FP \\
\midrule
Gemini 3.1 Pro        & 12\% & 70\% & 54\% & 33\% & 57\% & 17\% \\
Gemini 3.5 Flash      & 10\% & 57\% & 50\% & 34\% & 56\% & 12\% \\
Gemini 3.1 Flash-Lite & 27\% & 50\% & 22\% & 20\% & 39\% & 51\% \\
DeepSeek V4 Pro       & 14\% & 54\% & 30\% & 18\% & 37\% & 48\% \\
GLM 5.2               & 12\% & 68\% & 46\% & 34\% & 32\% & 23\% \\
Gemma 4 31B           & 23\% & 76\% & 25\% & 12\% & 42\% & 46\% \\
Qwen 3.6 27B          & 20\% & 58\% & 44\% & 31\% & 38\% & 39\% \\
Gemma 4 26B           & 26\% & 66\% &  0\% &  4\% & 24\% & 46\% \\
\midrule
Mean                  & 19\% & 61\% & 44\% & 28\% & 37\% & 44\% \\
\bottomrule
\end{tabular}
\end{table}

\subsection{Limitations of the Evaluation}\label{sec:limitations}
\paragraph{Balanced base rate.}
The dataset is balanced at 50\% malware, which eases per-class comparison but does not reflect operational intake streams where benign samples dominate.
Recall, specificity, and tBA are prevalence-independent, but escalation-queue precision is not: at a lower malware prevalence the same verdict behavior would yield queues with substantially lower precision.
The reported queue composition should therefore be read as a comparison between models, not a deployment estimate.

\paragraph{Evaluation labels and excluded gray zone.}
VirusTotal consensus supplies evaluation labels, not definitive ground truth.
At dataset-construction time, we required zero VT detections for benign and at least ten for malware. 
Samples with 1--9 detections were excluded, so we do not characterize binaries on which engines disagree, even though these may be especially relevant to escalation.

\paragraph{Scope of the comparison.}
The experiments compare LLMs and individual components of TRIAGE-CPG, not the complete workflow against trained static classifiers, commercial detectors, or dynamic-analysis systems.
This study addresses the prior question of whether current LLMs can implement a budgeted triage policy from partial static evidence at all, and its ternary verdicts, escalation queues, and behavioral reports have no direct analogue in a classifier score, so a fair comparison requires its own cost-matched design.
The results therefore do not establish that LLM-based triage is more accurate or cheaper than simpler alternatives.

\paragraph{Fixed protocol.}
All models use one configuration selected with Gemini 3.1 Flash-Lite on the development split.
The budget values are one operating point, not model-specific optima, and different prompts, budgets, or action interfaces could shift the results, though the shared configuration ensures each model receives the same evidence and actions.

\paragraph{Preprocessing and input scope.}
The evidence budget excludes decompilation and CPG construction, so we do not measure end-to-end analysis time.
The evaluation is conditional on producing a valid CPG, covers only Windows PE binaries, excludes .NET, and neither executes nor unpacks samples, so runtime-only behavior remains outside the workflow.
It also assumes non-adversarial inputs: instructions embedded in decompiled code can manipulate LLM analysis agents \cite{crawfordAutomaticallyAttackingSoftware2026}, and we do not evaluate the workflow under such attacks.
Finally, the model set reflects Vertex AI availability at experiment time, and serving-side updates may change individual results.

\section{Related Work}

\paragraph{Malware analysis.}
Automated malware analysis has been shaped by sample volume for two decades.
Behavior-based systems clustered thousands of daily sandbox reports so that analysts inspect representatives instead of every sample \cite{bayerScalableBehaviorBasedMalware}, and dynamic analysis remains the backbone of that pipeline \cite{egeleSurveyAutomatedDynamic2008}.
Static approaches face different obstacles.
Obfuscation can defeat static detection in principle \cite{moserLimitsStaticAnalysis2007}, runtime packing is pervasive among malicious samples \cite{ugarte-pedreroSoKDeepPacker2015}, and classifiers built on static features degrade once benign programs are packed as well \cite{aghakhaniWhenMalwarePackin2020}, which matches the packed-wrapper failure class of \Cref{sec:failure}.
A survey of practitioners reports that analysts use automated results as high-level guidance on where to look and to avoid over-inspecting individual samples \cite{botacinWhatMalwareAnalysts2024}.
Our study focuses on this bounded first pass and examines how LLMs use partial static-code evidence to clear, flag, or escalate a binary.

\paragraph{Code property graphs.}
Code property graphs merge the syntax, control flow, and data dependencies of a program into one queryable graph, introduced for vulnerability mining in C code \cite{yamaguchiModelingDiscoveringVulnerabilities2014} and later generalized into language-independent analysis platforms \cite{weissLanguageIndependentAnalysisPlatform2022, kuchlerRepresentingLLVMIRCode2022}.
Vulnerability discovery on decompiled binaries approaches source-level results in case studies \cite{mantovaniConvergenceSourceCode2022}, although decompiler evaluations show the recovered code carries semantic errors and template-like artifacts \cite{caoEvaluatingEffectivenessDecompilers2024}, noise that any consumer of a CPG inherits.
LLMxCPG uses CPG-guided slicing to fit vulnerability-relevant code into an LLM context \cite{lekssaysLLMxCPGContextAwareVulnerability}.
Our workflow instead uses the function-level call graph primarily as a navigation interface for selecting a bounded number of function reads during malware triage.

\paragraph{Graph signals and function prioritization.}
Scoring call-graph nodes with inexpensive structural signals has a long record in malware detection: PageRank has been used for malware categorization \cite{kangPageRankMalwareCategorization2015}, and MalScan classifies Android apps from centrality vectors of sensitive APIs over the call graph \cite{wuMalScanFastMarketWide2019}.
A second line of work ranks code units for human review rather than classifying whole samples: MKLDroid assigns malice scores down to basic blocks and methods \cite{narayananMultiviewContextawareApproach2018}, Genius ranks binary functions for manual confirmation effort \cite{fengScalableGraphbasedBug2016}, and DeepReflect localizes potentially malicious functionality to reduce reverse-engineering effort \cite{downingDeepReflectDiscoveringMalicious2021}.
In contrast, EPS is an untrained ranking over three static signals that gives every evaluated LLM the same initial function set, which we compare with several inexpensive alternatives without claiming optimality.

\paragraph{LLMs for code and malware analysis.}
LLMs increasingly support binary-code analysis, including decompilation \cite{tanLLM4DecompileDecompilingBinary2024, wangLargeLanguageModels2026}, malware summarization \cite{luMalsightExploringMalicious2025}, Android malware from code slices \cite{qianLAMDContextDrivenAndroid2025}, and malicious-code localization \cite{sunMalLocFinegrainedAndroid}.
Most closely, TTPDetect retrieves candidate functions and incrementally explores call-graph context to attribute ATT\&CK techniques in stripped malware binaries \cite{xuanIdentifyingAdversaryTactics2026}.
Our study addresses a different decision: given both benign and malicious binaries and a fixed function-read budget, how do different LLMs choose among clearing, flagging, and escalating a sample?
Recent work also demonstrates that instructions embedded in decompiled code can manipulate LLM-based reverse-engineering agents \cite{crawfordAutomaticallyAttackingSoftware2026}.

\paragraph{Triage with a reject option.}
Letting a classifier reject uncertain inputs goes back to Chow's error--reject tradeoff \cite{chowOptimumRecognitionError1970a}, developed further as selective classification \cite{geifmanSelectiveClassificationDeep2017}, learning to defer to a human expert \cite{mozannarConsistentEstimatorsLearning2020}, and cost-sensitive learning that prices the error directions independently \cite{elkanFoundationsCostsensitiveLearning2001}.
A separate budgeted line acquires costly features sequentially and stops once the evidence suffices \cite{jiCostsensitiveFeatureAcquisition2007}. 
Security operations already run on this logic, as alert-triage systems suppress or escalate alerts to protect analyst attention \cite{hassanNoDozeCombattingThreat2019}.
Our setting instantiates feature acquisition with function reads as the unit of cost and interprets rejection as escalation.

\section{Conclusion}
We studied how current LLMs triage a decompiled binary from a small, graph-selected subset of its functions, using TRIAGE-CPG as a common evidence interface combining LLM-free seed selection, bounded call-graph exploration, and a ternary verdict.
Across eight models under one protocol, the highest-scoring model reaches a strict balanced accuracy of 0.805 and a tBA of 0.855 while inspecting about 40 of a mean \num{1000} recovered functions per binary.
The larger finding is that models occupy substantially different recall--specificity and escalation operating points, spanning both closed-source and open-weight families, so model choice itself determines the triage policy. These results characterize LLM behavior under limited static evidence rather than establishing superiority over conventional malware detectors.
Future work should benchmark the complete workflow against non-LLM detection baselines and connect the escalation queue with unpacking or dynamic analysis, where wrapped samples concentrate.


\bibliographystyle{ACM-Reference-Format}
\bibliography{references}

@string{oakland = {{IEEE} Symposium on Security and Privacy ({SP})}}

@string{ccs     = {{ACM} Conference on Computer and Communications Security ({CCS})}}

@string{usenix  = {{USENIX} Security Symposium}}

@string{ndss    = {Network and Distributed System Security Symposium ({NDSS})}}

@string{acsac   = {Annual Computer Security Applications Conference ({ACSAC})}}

@string{raid    = {International Symposium on Research in Attacks, Intrusions and Defenses ({RAID})}}

@string{asiaccs = {{ACM} Asia Conference on Computer and Communications Security ({AsiaCCS})}}

@string{isc     = {Information Security Conference ({ISC})}}

@string{ase     = {{IEEE/ACM} International Conference on Automated Software Engineering ({ASE})}}

@string{issta   = {{ACM} International Symposium on Software Testing and Analysis ({ISSTA})}}

@string{nips    = {Advances in Neural Information Processing Systems ({NeurIPS})}}

@string{iclr    = {International Conference on Learning Representations ({ICLR})}}

@string{cnsse   = {International Conference on Computer Network Security and Software Engineering ({CNSSE})}}

@string{kdd     = {{ACM} {SIGKDD} Conference on Knowledge Discovery and Data Mining ({KDD})}}

@string{racs    = {Conference on Research in Adaptive and Convergent Systems ({RACS})}}

@string{emnlp   = {Conference on Empirical Methods in Natural Language Processing ({EMNLP})}}

@string{spw     = {{IEEE} Security and Privacy Workshops ({SPW})}}

@string{icsme   = {{IEEE} International Conference on Software Maintenance and Evolution ({ICSME})}}

@string{icml    = {International Conference on Machine Learning ({ICML})}}

@string{ijcai   = {International Joint Conference on Artificial Intelligence ({IJCAI})}}

@inbook{baileyAutomatedClassificationAnalysis2007,
    author = {Bailey, Michael and Oberheide, Jon and Andersen, Jon and Mao, Z. Morley and Jahanian, Farnam and Nazario, Jose},
    title = {Automated {Classification} and {Analysis} of {Internet} {Malware}},
    booktitle = {Recent Advances in Intrusion Detection},
    year = {2007},
    pages = {178--197},
    doi = {10.1007/978-3-540-74320-0_10},
}

@inproceedings{bayerScalableBehaviorBasedMalware,
    author = {Bayer, Ulrich and Milani Comparetti, Paolo and Hlauschek, Clemens and Kruegel, Christopher and Kirda, Engin},
    title = {Scalable, {Behavior}-{Based} {Malware} {Clustering}},
    booktitle = ndss,
    year = {2009},
}

@article{egeleSurveyAutomatedDynamic2008,
    author = {Egele, Manuel and Scholte, Theodoor and Kirda, Engin and Kruegel, Christopher},
    title = {A Survey on Automated Dynamic Malware-Analysis Techniques and Tools},
    journal = {ACM Computing Surveys},
    year = {2012},
    volume = {44},
    number = {2},
    pages = {6:1--6:42},
    doi = {10.1145/2089125.2089126},
}

@inproceedings{botacinWhatMalwareAnalysts2024,
    author = {Botacin, Marcus},
    title = {What Do Malware Analysts Want from Academia? {A} Survey on the State of the Practice to Guide Research Developments},
    booktitle = raid,
    year = {2024},
}

@inproceedings{grazianoNeedlesHaystackMining2015,
    author = {Graziano, Mariano and Canali, Davide and Bilge, Leyla and Lanzi, Andrea and Balzarotti, Davide},
    title = {Needles in a {Haystack}: Mining Information from Public Dynamic Analysis Sandboxes for Malware Intelligence},
    booktitle = usenix,
    year = {2015},
}

@inproceedings{kiratBareCloudBaremetalAnalysisbased,
    author = {Kirat, Dhilung and Vigna, Giovanni and Kruegel, Christopher},
    title = {{BareCloud}: Bare-Metal Analysis-Based Evasive Malware Detection},
    booktitle = usenix,
    year = {2014},
}

@inproceedings{miramirkhaniSpotlessSandboxesEvading2017,
    author = {Miramirkhani, Najmeh and Appini, Mahathi Priya and Nikiforakis, Nick and Polychronakis, Michalis},
    title = {Spotless Sandboxes: Evading Malware Analysis Systems Using Wear-and-Tear Artifacts},
    booktitle = oakland,
    year = {2017},
}

@inproceedings{kolbitschPowerProcrastinationDetection2011,
    author = {Kolbitsch, Clemens and Kirda, Engin and Kruegel, Christopher},
    title = {The Power of Procrastination: Detection and Mitigation of Execution-Stalling Malicious Code},
    booktitle = ccs,
    year = {2011},
}

@inproceedings{DoesEverySecond,
    author = {K{"u}chler, Alexander and Mantovani, Alessandro and Han, Yufei and Bilge, Leyla and Balzarotti, Davide},
    title = {Does Every Second Count? Time-Based Evolution of Malware Behavior in Sandboxes},
    booktitle = ndss,
    year = {2021},
}

@inproceedings{moserLimitsStaticAnalysis2007,
    author = {Moser, Andreas and Kruegel, Christopher and Kirda, Engin},
    title = {Limits of Static Analysis for Malware Detection},
    booktitle = acsac,
    year = {2007},
}

@inproceedings{marcelliHowMachineLearning2022,
    author = {Marcelli, Andrea and Graziano, Mariano and Ugarte{-}Pedrero, Xabier and Fratantonio, Yanick and Mansouri, Mohamad and Balzarotti, Davide},
    title = {How Machine Learning Is Solving the Binary Function Similarity Problem},
    booktitle = usenix,
    year = {2022},
}

@inproceedings{yangSWEagentAgentcomputerInterfaces2024,
    author = {Yang, John and Jimenez, Carlos E. and Wettig, Alexander and Lieret, Kilian and Yao, Shunyu and Narasimhan, Karthik and Press, Ofir},
    title = {{SWE}-agent: Agent-Computer Interfaces Enable Automated Software Engineering},
    booktitle = nips,
    year = {2024},
}

@inproceedings{yao2023react,
    author = {Yao, Shunyu and Zhao, Jeffrey and Yu, Dian and Du, Nan and Shafran, Izhak and Narasimhan, Karthik R. and Cao, Yuan},
    title = {{ReAct}: Synergizing Reasoning and Acting in Language Models},
    booktitle = iclr,
    year = {2023},
}

@inproceedings{yamaguchiModelingDiscoveringVulnerabilities2014,
    author = {Yamaguchi, Fabian and Golde, Nico and Arp, Daniel and Rieck, Konrad},
    title = {Modeling and Discovering Vulnerabilities with Code Property Graphs},
    booktitle = oakland,
    year = {2014},
}

@misc{weissLanguageIndependentAnalysisPlatform2022,
    author = {Weiss, Konrad and Banse, Christian},
    title = {A Language-Independent Analysis Platform for Source Code},
    year = {2022},
    note = {arXiv:2203.08424; preprint},
    doi = {10.48550/arXiv.2203.08424},
}

@inproceedings{wangLargeLanguageModels2026,
    author = {Wang, Yuanchao and Tian, Ye and Wei, Hongmei and Wang, Fei},
    title = {Large Language Models for Binary Decompilation: Paradigms, Evaluation, and Challenges},
    booktitle = cnsse,
    year = {2026},
}

@inproceedings{mohammadiEvaluationBenchmarkingLLM2025,
    author = {Mohammadi, Mahmoud and Li, Yipeng and Lo, Jane and Yip, Wendy},
    title = {Evaluation and Benchmarking of {LLM} Agents: {A} Survey},
    booktitle = kdd,
    year = {2025},
}

@inproceedings{lekssaysLLMxCPGContextAwareVulnerability,
    author = {Lekssays, Ahmed and Mouhcine, Hamza and Tran, Khang and Yu, Ting and Khalil, Issa},
    title = {{LLMxCPG}: Context-Aware Vulnerability Detection Through Code Property Graph-Guided Large Language Models},
    booktitle = usenix,
    year = {2025},
}

@article{chowOptimumRecognitionError1970a,
    author = {Chow, C. K.},
    title = {On Optimum Recognition Error and Reject Tradeoff},
    journal = {IEEE Transactions on Information Theory},
    year = {1970},
    volume = {16},
    number = {1},
    pages = {41--46},
    doi = {10.1109/TIT.1970.1054406},
}

@misc{horsHorsicqDetectItEasy2026,
    author = {{horsicq}},
    title = {{Detect It Easy (DiE)}},
    year = {2014},
    url = {https://github.com/horsicq/Detect-It-Easy},
    note = {accessed 2026-07-09},
}

@inproceedings{ugarte-pedreroSoKDeepPacker2015,
    author = {Ugarte{-}Pedrero, Xabier and Balzarotti, Davide and Santos, Igor and Bringas, Pablo G.},
    title = {{SoK}: Deep Packer Inspection: {A} Longitudinal Study of the Complexity of Run-Time Packers},
    booktitle = oakland,
    year = {2015},
}

@inproceedings{aghakhaniWhenMalwarePackin2020,
    author = {Aghakhani, Hojjat and Gritti, Fabio and Mecca, Francesco and Lindorfer, Martina and Ortolani, Stefano and Balzarotti, Davide and Vigna, Giovanni and Kruegel, Christopher},
    title = {When Malware Is Packin' Heat; Limits of Machine Learning Classifiers Based on Static Analysis Features},
    booktitle = ndss,
    year = {2020},
}

@inbook{kuchlerRepresentingLLVMIRCode2022,
    author = {K{"u}chler, Alexander and Banse, Christian},
    title = {Representing {LLVM-IR} in a Code Property Graph},
    booktitle = isc,
    year = {2022},
    pages = {360--380},
    doi = {10.1007/978-3-031-22390-7_21},
}

@inproceedings{mantovaniConvergenceSourceCode2022,
    author = {Mantovani, Alessandro and Compagna, Luca and Shoshitaishvili, Yan and Balzarotti, Davide},
    title = {The Convergence of Source Code and Binary Vulnerability Discovery---A Case Study},
    booktitle = asiaccs,
    year = {2022},
}

@inproceedings{caoEvaluatingEffectivenessDecompilers2024,
    author = {Cao, Ying and Zhang, Runze and Liang, Ruigang and Chen, Kai},
    title = {Evaluating the Effectiveness of Decompilers},
    booktitle = issta,
    year = {2024},
}

@inproceedings{kangPageRankMalwareCategorization2015,
    author = {Kang, BooJoong and Yerima, Suleiman and McLaughlin, Kieran and Sezer, Sakir},
    title = {{PageRank} in Malware Categorization},
    booktitle = racs,
    year = {2015},
}

@inproceedings{wuMalScanFastMarketWide2019,
    author = {Wu, Yueming and Li, Xiaodi and Zou, Deqing and Yang, Wei and Zhang, Xin and Jin, Hai},
    title = {{MalScan}: Fast Market-Wide Mobile Malware Scanning by Social-Network Centrality Analysis},
    booktitle = ase,
    year = {2019},
}

@article{narayananMultiviewContextawareApproach2018,
    author = {Narayanan, Annamalai and Chandramohan, Mahinthan and Chen, Lihui and Liu, Yang},
    title = {A Multi-View Context-Aware Approach to Android Malware Detection and Malicious Code Localization},
    journal = {Empirical Software Engineering},
    year = {2018},
    volume = {23},
    number = {3},
    pages = {1222--1274},
    doi = {10.1007/s10664-017-9539-8},
}

@inproceedings{fengScalableGraphbasedBug2016,
    author = {Feng, Qian and Zhou, Rundong and Xu, Chengcheng and Cheng, Yao and Testa, Brian and Yin, Heng},
    title = {Scalable Graph-Based Bug Search for Firmware Images},
    booktitle = ccs,
    year = {2016},
}

@inproceedings{downingDeepReflectDiscoveringMalicious2021,
    author = {Downing, Evan and Mirsky, Yisroel and Park, Kyuhong and Lee, Wenke},
    title = {{DeepReflect}: Discovering Malicious Functionality Through Binary Reconstruction},
    booktitle = usenix,
    year = {2021},
}

@inproceedings{tanLLM4DecompileDecompilingBinary2024,
    author = {Tan, Hanzhuo and Luo, Qi and Li, Jing and Zhang, Yuqun},
    title = {{LLM4Decompile}: Decompiling Binary Code with Large Language Models},
    booktitle = emnlp,
    year = {2024},
}

@article{luMalsightExploringMalicious2025,
    author = {Lu, Haolang and Peng, Hongrui and Nan, Guoshun and Cui, Jiaoyang and Wang, Cheng and Jin, Weifei and Wang, Songtao and Pan, Shengli and Tao, Xiaofeng},
    title = {{Malsight}: Exploring Malicious Source Code and Benign Pseudocode for Iterative Binary Malware Summarization},
    journal = {IEEE Transactions on Information Forensics and Security},
    year = {2025},
    volume = {20},
    pages = {6733--6747},
    doi = {10.1109/TIFS.2025.3583552},
}

@inproceedings{qianLAMDContextDrivenAndroid2025,
    author = {Qian, Xingzhi and Zheng, Xinran and He, Yiling and Yang, Shuo and Cavallaro, Lorenzo},
    title = {{LAMD}: Context-Driven Android Malware Detection and Classification with {LLMs}},
    booktitle = spw,
    year = {2025},
}

@inproceedings{sunMalLocFinegrainedAndroid,
    author = {Sun, Tiezhu and Alecci, Marco and Pilgun, Aleksandr and Song, Yewei and Tang, Xunzhu and Samhi, Jordan and Bissyand{\'e}, Tegawend{\'e} F. and Klein, Jacques},
    title = {{MalLoc}: Toward Fine-Grained Android Malicious Payload Localization via {LLMs}},
    booktitle = icsme,
    year = {2025},
}

@misc{xuanIdentifyingAdversaryTactics2026,
    author = {Xuan, Zhou and Xu, Xiangzhe and Zheng, Mingwei and Tan, Louis Zheng-Hua and Guo, Jinyao and Zhang, Tiantai and Yu, Le and Wang, Chengpeng and Zhang, Xiangyu},
    title = {Identifying Adversary Tactics and Techniques in Malware Binaries with an {LLM} Agent},
    year = {2026},
    note = {arXiv:2602.06325; preprint},
    doi = {10.48550/arXiv.2602.06325},
}

@article{crawfordAutomaticallyAttackingSoftware2026,
    author = {Crawford, Brian and Phillips, Justin and McClure, Patrick},
    title = {Automatically Attacking Software Reverse Engineering {AI} Agents},
    journal = {European Conference on Cyber Warfare and Security},
    year = {2026},
    volume = {25},
    number = {1},
    pages = {199--205},
    doi = {10.34190/eccws.25.1.4597},
}

@inproceedings{geifmanSelectiveClassificationDeep2017,
    author = {Geifman, Yonatan and El-Yaniv, Ran},
    title = {Selective Classification for Deep Neural Networks},
    booktitle = nips,
    year = {2017},
}

@inproceedings{mozannarConsistentEstimatorsLearning2020,
    author = {Mozannar, Hussein and Sontag, David},
    title = {Consistent Estimators for Learning to Defer to an Expert},
    booktitle = icml,
    year = {2020},
}

@inproceedings{elkanFoundationsCostsensitiveLearning2001,
    author = {Elkan, Charles},
    title = {The Foundations of Cost-Sensitive Learning},
    booktitle = ijcai,
    year = {2001},
}

@article{jiCostsensitiveFeatureAcquisition2007,
    author = {Ji, Shihao and Carin, Lawrence},
    title = {Cost-Sensitive Feature Acquisition and Classification},
    journal = {Pattern Recognition},
    year = {2007},
    volume = {40},
    number = {5},
    pages = {1474--1485},
    doi = {10.1016/j.patcog.2006.11.008},
}

@inproceedings{hassanNoDozeCombattingThreat2019,
    author = {Hassan, Wajih Ul and Guo, Shengjian and Li, Ding and Chen, Zhengzhang and Jee, Kangkook and Li, Zhichun and Bates, Adam},
    title = {{NoDoze}: Combatting Threat Alert Fatigue with Automated Provenance Triage},
    booktitle = ndss,
    year = {2019},
}

\appendix 

\section{Function Selection and EPS Weights}\label{app:func-select}
\Cref{tab:seed-size} reports the fixed-seed-count sweep across binary-size classes that motivates the proportional seeding rule of \Cref{sec:funcprior}.

\begin{table}[!ht]
\centering
\small
\setlength{\tabcolsep}{3pt}
\caption{Development-split balanced accuracy by binary size and fixed seed count (exploration off). Bold marks each size class's best count.}
\label{tab:seed-size}
\begin{tabular}{lrrrrrrr}
\toprule
Binary size & $n$ & \multicolumn{6}{c}{Seeds per binary} \\
\cmidrule(lr){3-8}
 & & 10 & 25 & 50 & 75 & 100 & 150 \\
\midrule
Small ($<$100)  & 63  & \textbf{0.819} & 0.795 & 0.732 & 0.766 & 0.787 & 0.787 \\
Mid (100--750)  & 68  & 0.560 & \textbf{0.821} & 0.815 & 0.778 & 0.741 & 0.692 \\
Large ($>$750)  & 69  & 0.639 & 0.723 & 0.712 & 0.686 & \textbf{0.730} & 0.670 \\
\midrule
All             & 200 & 0.690 & \textbf{0.810} & 0.775 & 0.755 & 0.750 & 0.720 \\
\bottomrule
\end{tabular}
\end{table}

\paragraph{EPS weight selection.}\label{app:eps-grid}
We sweep all 70 EPS weight vectors with components drawn from $\{0, 0.25, 0.5, 0.75, 1\}$ summing to one on the train split, scored under the strict verdict metric.
The best vector places weight 0.50 on the boundary flag ($B$) and 0.25 each on reverse PageRank and code length, reaching 0.720 balanced accuracy and 0.756 F1, with the three runner-up configurations scoring between 0.690 and 0.710.
The deployed vector assigns zero weight to forward PageRank and the API-rarity flag.
We deploy the train-best vector and freeze it for every later experiment.

\section{Exploration}\label{app:exploration}
\Cref{tab:exploration-detail} gives the per-configuration detail behind the exploration experiments of \Cref{sec:eval-exploration}: the paired on/off comparison, agent-guided against random targeting, and the seed selectors rerun under the full pipeline.
For the matched-action random control in panel (b), we set the exploration cap to 13, rounding the agent's mean of 13.2 exploration actions. Because agent actions may query the graph while random actions always analyze a function, matching actions does not imply matching function reads (5.2 vs.\ 11.3).

\begin{table}[!ht]
\centering
\small
\setlength{\tabcolsep}{4pt}
\caption{Development-split exploration analyses under the strict verdict. 
(a) Net benign/malware binaries corrected minus hurt by enabling exploration across nine seed/reporting configurations; bold marks the deployed configuration. 
(b) Agent-guided and random targeting under full and action-matched budgets, reporting realized function reads and balanced-accuracy ranges. 
(c) Seed-selector performance under the full pipeline.}
\label{tab:exploration-detail}

\begin{subtable}{\linewidth}
\caption{Exploration on vs.\ off (benign net / malware net)}
\label{tab:exploration-onoff}
\centering
\begin{tabular}{lrrr}
\toprule
Reporting & \multicolumn{3}{c}{Seeds per binary} \\
\cmidrule(lr){2-4}
budget & 25 & 75 & 150 \\
\midrule
10 & 5 / $-3$ & 6 / $-2$ & 13 / $-6$ \\
20 & 4 / $-1$ & \textbf{10 / 0} & 10 / $-3$ \\
30 & 1 / $-4$ & 5 / $-2$ & 14 / $-2$ \\
\bottomrule
\end{tabular}
\end{subtable}\par\medskip

\begin{subtable}{\linewidth}
\caption{Exploration policy}
\label{tab:exploration-policy}
\centering
\begin{tabular}{lrrr}
\toprule
Policy & Arms & Reads & Bal.\ Acc. \\
\midrule
Agent-guided           & 5 & 5.2  & 0.800--0.820 \\
Random, full budget    & 3 & 25.4 & 0.775--0.805 \\
Random, action-matched & 3 & 11.3 & 0.760--0.815 \\
\bottomrule
\end{tabular}
\end{subtable}\par\medskip

\begin{subtable}{\linewidth}
\caption{Seed selector, full pipeline}
\label{tab:exploration-selectors}
\centering
\begin{tabular}{lrrr}
\toprule
Selector & Recall & Spec. & Bal.\ Acc. \\
\midrule
EPS          & 0.890 & 0.710 & 0.800 \\
Largest code & 0.820 & 0.700 & 0.760 \\
Random       & 0.870 & 0.720 & 0.795 \\
\bottomrule
\end{tabular}
\end{subtable}\par
\end{table}

\section{Open Science}
The accompanying artifact contains the TRIAGE-CPG implementation, benchmark manifests with SHA-256 hashes, labels, and split assignments for all \num{1000} binaries, frozen experiment configurations, aggregate artifacts underlying all reported analyses, sanitized per-sample behavioral reports and VirusTotal-derived judgments for the eight locked test runs, and the corresponding analysis scripts.
The artifact is available at \url{https://doi.org/10.5281/zenodo.22771897}.
The binaries originate from VirusTotal and cannot be redistributed under its terms of service.
We publish their hashes so that VirusTotal subscribers can retrieve the same samples, subject to access and continued availability, together with the CPG extraction configuration required to reconstruct the graphs.

\section{Ethical Considerations}
This work analyzes existing Windows malware obtained from VirusTotal and creates no new malware, exploit code, or attack tooling.
The pipeline never executes the samples, and the binaries are not redistributed.
The generated reports describe behavior at the level of standard threat-intelligence write-ups and omit decompiled function bodies and instructions for building or improving malware.
The intended and evaluated use is defensive, reducing the triage load of security analysts, and the paper documents the failure modes an operator must account for before relying on the system.
No human subjects or personal data are involved.

\section*{Use of Generative AI}
Pi (\texttt{pi.dev}), an agentic coding harness, was used with on-premises language models and models provided by Anthropic and OpenAI to assist with developing parts of the experimental code and analysis scripts and with editing and polishing the manuscript.
The authors manually reviewed all AI-assisted code and text before inclusion.
The research questions, system design, experimental protocol, and interpretation of the results were determined by the authors, who take full responsibility for the paper.

\end{document}